\documentclass[a4paper,fleqn]{cas-sc}

\usepackage[numbers]{natbib}

\def\tsc#1{\csdef{#1}{\textsc{\lowercase{#1}}\xspace}}
\tsc{WGM}
\tsc{QE}
\tsc{EP}
\tsc{PMS}
\tsc{BEC}
\tsc{DE}

\usepackage{booktabs} \usepackage{longtable} \usepackage{array} \usepackage{ragged2e} \newcolumntype{L}[1]{>{\RaggedRight\arraybackslash}p{#1}}

\usepackage[table,xcdraw,dvipsnames]{xcolor}
\definecolor{myblue}{RGB}{0,0,255}
\newcommand{\BlueText}[1]{{\color{black}#1}}

\usepackage{tabularx}
\usepackage{float}

\usepackage{booktabs}
\usepackage{wasysym}



\makeatletter
\AtEndDocument{\clearpage\immediate\write\@auxout{\string\csxdef{lastpage}{\the\numexpr\value{page}-1\relax}}}
\makeatother

\begin{document}
\let\WriteBookmarks\relax
\def\floatpagepagefraction{1}
\def\textpagefraction{.001}

\shorttitle{Multi-agent LLMs on the Cloud-Edge-IoT Continuum}

\title [mode = title]{Multi-agent LLMs on the Cloud-Edge-IoT Continuum: A Systematic Mapping Study on Architectures, Deployment, and Evaluation}     
\shortauthors{M.~T. Moghaddam et~al.}

\author[1]{Mahyar Tourchi Moghaddam}[orcid=0000-0001-5028-7546]
\cormark[1]
\ead{mtmo@mmmi.sdu.dk}

\affiliation[1]{organization={University of Southern Denmark},
    city={Odense},
    country={Denmark}}

\author[1]{Tiziano Santilli}[orcid=0009-0005-9505-3846]
\ead{tisa@mmmi.sdu.dk}

\author[1]{Leo Wilhelm Lierse}[orcid=0009-0008-3232-4954]
\ead{leoli@mmmi.sdu.dk}

\author[2]{Sebastian Werner}[orcid=0000-0001-8051-7226]
\ead{werner.sebastian@tum.de}

\affiliation[2]{organization={Technical University of Munich},
    city={Munich},
    country={Germany}}

\author[1]{Francesco Daghero}[orcid=0000-0001-9595-7216]
\ead{fdag@mmmi.sdu.dk}

\cortext[1]{Corresponding author.}

\begin{abstract}
Large language model (LLM) applications are increasingly built as systems of cooperating components rather than as single prompt-response pipelines. Internet-of-Things (IoT) applications add constraints of their own, because sensing, reasoning, control, and actuation are spread across devices, edge nodes, and cloud services to meet latency, privacy, and bandwidth requirements. How multi-agent LLM systems are organized and deployed across this cloud-edge-IoT continuum has not been mapped systematically. We therefore conducted a systematic mapping study. From 831 records retrieved up to March 2026, we selected 45 primary studies and classified each one by architecture family, deployment pattern, application domain, and evaluation practice. We also recorded how each system was evaluated and whether its code or data are public. Five architecture families recur in the corpus. Specialized agent teams run next to sensors, planners delegate subtasks to workers, and small local models cooperate with large cloud models. A fourth group federates model adaptation across devices, and a large fifth group builds the supporting infrastructure for offloading, communication, memory, and serving. Deployment relies on six recurring patterns, ranging from local-only execution to control loops gated by policy checks or human approval. Twenty-four of the 45 studies target IoT, industrial, smart-home, healthcare, vehicular, or robotics settings, and 29 evaluate on physical hardware, where open models below 8B parameters dominate. The corpus also has clear weaknesses. Only six studies release code or data, and only 15 of the 45 report ablation experiments. Privacy motivates many local designs but is directly measured in only two studies. Latency, task quality, cost, privacy, and resilience are rarely reported together. Based on the mapping, we list the reporting items needed to make future evaluations comparable, give deployment guidance for common IoT constraints, and identify open problems. The protocol, screening decisions, and coded data are publicly archived.
\end{abstract}

\begin{highlights}
\item Maps 45 studies of multi-agent LLM systems across cloud, edge, and IoT
\item Finds five recurring architecture families and six deployment patterns
\item 29 of 45 studies evaluate on physical hardware; small open models lead the edge
\item Six of 45 studies share artifacts, and evaluation setups rarely overlap
\item Gives a reporting checklist and deployment guidance for IoT settings
\end{highlights}

\begin{keywords}
LLM \sep Agentic AI \sep Cloud \sep Edge Computing \sep IoT \sep Cloud-edge Continuum \sep Edge AI \sep Architecture \sep Deployment \sep Benchmark
\end{keywords}

\maketitle

\section{Introduction}

Large Language Model (LLM) systems are increasingly moving from monolithic prompt-response pipelines to {\bf \em agentic systems} that combine reasoning, action, tool use, interaction with external environments, and memory \cite{agentic}. 
\BlueText{A multi-agent LLM system, as used in this paper, is a software system in which several language-model-based components cooperate on a task. One component may interpret intent, while others retrieve context, call tools, or verify and coordinate execution.}
Researchers investigated agentic patterns by explicitly interleaving reasoning traces and actions to reduce hallucinations and increase interpretability on interactive tasks \cite{react}. Others made multi-agent conversation programmable through a software architecture for task decomposition and collaboration \cite{autogen}.
The community also investigated the shift from single-agent pipelines to multi-agent LLM systems, in which specialized agents coordinate through dialogue, role assignment, debate, or workflow graphs \cite{single}. 
Role-playing and role specialization help explore scalable cooperation by assigning roles and structuring {\em societies} of agents \cite{role}. In application-rich domains, multi-agent decomposition has been studied for software development \cite{chatdev} and Standard Operating Procedure (SOP)-like workflows \cite{metagpt}, both emphasizing role specialization, structured interaction, and intermediate verification to mitigate cascading errors.

In parallel, \BlueText{modern IoT systems rarely place all intelligence in a single place.} 
Therefore, the computing substrate is shifting from cloud-only to {\bf \em edge-cloud-IoT} continuum deployments \cite{continuum}. Edge computing is motivated by faster response (latency), bandwidth savings, privacy, offline reliability, and local processing near data sources. Fog and continuum concepts formalize multi-tier architectures spanning devices, gateways, and cloud resources, providing a vocabulary for {\em where computation happens} and why multi-layer placements matter.

Current deployments combine the two developments. IoT devices and sensors, edge gateways, near-edge or multi-access edge tiers, and cloud services work together. 
Multi-agent work typically emphasizes orchestration, role specialization, interaction patterns, and coordination among LLM-based agents, while edge LLM and small language model (SLM) work tends to focus on efficient deployment of individual models through small-model design, compression, and inference optimization rather than distributed agent topologies and team protocols \cite{slm}.
However, real deployments increasingly combine both, as data is generated at the network edge, edge and near-edge resources provide low-latency regional processing, and cloud resources remain important for heavier computation and broader coordination \cite{edge}.

Despite the rapid growth of LLM-based multi-agent systems \cite{S1} and edge LLMs \cite{S2,zheng2025edgellm}, the intersection still lacks a continuum-aware architectural map that explains how roles, models, tools, memory, policies, and migration strategies are actually distributed across cloud, edge, and IoT layers.
\BlueText{We ask how these systems are organized, how they are deployed across device, edge, and cloud layers, which IoT problems they address, and how they are evaluated. These questions matter for IoT researchers and practitioners because the {\em agentic} label can hide very different decisions about locality, adaptation, state management, tool access, and human oversight.}

This study addresses four research questions:

\noindent{\bf \em RQ1.}
Which architectural families recur in cloud-edge-IoT multi-agent or agentic LLM systems?

\noindent{\bf \em RQ2.}
How is deployment realized, including role placement, offloading, synchronization, fallback, and migration?

\noindent{\bf \em RQ3.}
How are cloud-edge-IoT multi-agent LLM systems evaluated in practice, and what evaluation gaps exist?

\noindent{\bf \em RQ4.}
What open challenges remain for benchmarking, state management, security, lifecycle adaptation, oversight, and reproducibility?

We answer these questions with a systematic mapping study (SMS). From 831 candidate records, we selected 45 primary studies, extracted structured data from each, and classified them along shared dimensions.
\BlueText{Appendix~\ref{app:primary-studies} (Table~\ref{tab:primary-studies}) lists the 45 primary studies with their identifiers, first author and year, title, application domain, and venue. The identifiers P1 to P45 are citation labels. The primary studies are listed in a separate, second part of the reference list and are cited as \cite{P1} to \cite{P45} throughout the text, which keeps them distinguishable from the general references.}

This paper provides the following contributions:

\begin{enumerate}
    \item A taxonomy of the architectures and deployment patterns used by multi-agent LLM systems on the cloud-edge-IoT continuum.
    \item A mapping from deployment patterns to the quality attributes they serve.
    \item A study-by-study record of how each system was evaluated (conceptual argument, simulation or dataset experiments, physical hardware, or production data), whether baselines and ablations are reported, and whether code or data are public.
    \item An explicit account of how directly each study is tied to IoT and cyber-physical settings.
    \item A reporting checklist for comparable evaluations, deployment guidance for common IoT constraints, and a list of open problems.
\end{enumerate}

We provide the \textbf{replication package} used in this study at the following link\footnote{\url{https://doi.org/10.5281/zenodo.20631424}}.

\vspace{1em}
\BlueText{
For readers unfamiliar with certain technical terms and acronyms (e.g., LLM, SLM, offloading, federated adaptation, governance), Appendix \ref{app:glossary} provides a complete glossary of terms used in this paper.
}

\section{Related Surveys and Positioning}
\label{sec:related}

Several secondary studies border on our scope. Table~\ref{tab:related-surveys} compares this study with ten of them along five questions. Is the review systematic, with a documented protocol? Does it treat multi-agent LLM organization as a main subject? Does it treat cloud-edge-IoT deployment as a main subject? Does it analyze how the primary studies evaluate their systems? Does it provide a replication package?

Surveys of LLM-based multi-agent systems describe coordination mechanisms, communication topologies, and application areas \cite{S1,single,agentic}, and recent work separates ``AI agents'' from ``agentic AI'' at the level of definitions \cite{sapkota2025taxonomy}. These surveys take the computing infrastructure as given. They do not examine where roles, models, memory, and tools are placed across device, edge, and cloud tiers, or which mechanisms realize those placements. Surveys of edge LLMs and mobile edge intelligence cover model compression, on-device adaptation, caching, split inference, and serving in depth \cite{slm,S2,zheng2025edgellm,qu2025mei,friha2024llmedge}, but their unit of analysis is a single model rather than a set of cooperating agents. Surveys of the IoT-edge-cloud continuum \cite{continuum} describe the infrastructure without addressing LLM-based agents.

This study covers the intersection that these three groups leave open. To the best of our knowledge, it is the first systematic mapping of multi-agent LLM systems across the cloud-edge-IoT continuum. It differs from the narrative surveys above in four ways that a reader can check. The protocol, search strings, screening decisions, and exclusion reasons are archived in a public replication package. Every primary study is coded on four dimensions (architecture, deployment, domain, evaluation), so organizational form and runtime placement can be compared separately. For every study, we record how it was evaluated, whether baselines and ablations are reported, and whether artifacts are available. Finally, the mapping is used to derive a reporting checklist and deployment guidance rather than only a catalog. None of the ten studies in Table~\ref{tab:related-surveys} combines these properties.

\begin{table}[t]
\centering
\caption{Comparison with related surveys and reviews. SR = systematic review or mapping with a documented protocol; MA = multi-agent LLM organization is a main subject; CE = cloud-edge-IoT deployment is a main subject; EV = evaluation practice of the primary studies is analyzed; RP = public replication package. \CIRCLE\ = yes, \LEFTcircle\ = partly, \Circle\ = no.}
\label{tab:related-surveys}
\scriptsize
\begin{tabular}{p{0.44\linewidth}ccccc}
\toprule
\textbf{Study} & \textbf{SR} & \textbf{MA} & \textbf{CE} & \textbf{EV} & \textbf{RP}\\
\midrule
Guo et al.\ 2024, LLM-based multi-agents \cite{S1} & \Circle & \CIRCLE & \Circle & \LEFTcircle & \Circle\\
Li et al.\ 2024, LLM-MAS workflow/infrastructure \cite{single} & \Circle & \CIRCLE & \Circle & \Circle & \Circle\\
Ferrag et al.\ 2025, LLM reasoning to agents \cite{agentic} & \Circle & \CIRCLE & \Circle & \Circle & \Circle\\
Sapkota et al.\ 2025, agents vs.\ agentic AI \cite{sapkota2025taxonomy} & \Circle & \CIRCLE & \Circle & \Circle & \Circle\\
Wang et al.\ 2025, edge-efficient LLMs \cite{slm} & \Circle & \Circle & \CIRCLE & \Circle & \Circle\\
Zheng et al.\ 2025, edge LLM design/execution \cite{zheng2025edgellm} & \Circle & \Circle & \CIRCLE & \Circle & \Circle\\
Jiang et al.\ 2026, edge LLMs \cite{S2} & \Circle & \Circle & \CIRCLE & \Circle & \Circle\\
Qu et al.\ 2025, mobile edge intelligence for LLMs \cite{qu2025mei} & \Circle & \Circle & \CIRCLE & \Circle & \Circle\\
Friha et al.\ 2024, LLM-based edge intelligence \cite{friha2024llmedge} & \Circle & \LEFTcircle & \CIRCLE & \Circle & \Circle\\
Gkonis et al.\ 2023, IoT-edge-cloud continuum \cite{continuum} & \Circle & \Circle & \CIRCLE & \Circle & \Circle\\
\midrule
\textbf{This study} & \CIRCLE & \CIRCLE & \CIRCLE & \CIRCLE & \CIRCLE\\
\bottomrule
\end{tabular}
\end{table}

\section{Research Methodology}
\label{sec:method}

This study aims to define a classification scheme, structure the field, and identify concentrations and gaps in the literature on multi-agent LLMs across the computing continuum. 
The review follows established software engineering guidance for mapping studies and systematic literature reviews \cite{kitchenham2013systematic,kitchenham2007guidelines,wohlin2014guidelines,kitchenham2015evidence} and the reporting style follows PRISMA 2020 \cite{page2021prisma}
to support transparency, traceability, and reproducibility. The protocol has four phases: {\em i)} database search, {\em ii)} staged selection, {\em iii)} structured data extraction, and {\em iv)} coding and analysis.
To reduce selection bias and improve coverage, the protocol integrates backward and forward snowballing from key primary studies \cite{wohlin2014guidelines}.
We query major scientific indexing {\em databases} with strong coverage in software engineering, AI systems, and distributed systems, i.e., IEEE Xplore, ACM Digital Library, Scopus, and Web of Science, searched on full text. 
These databases are selected to balance recall, precision, and reproducibility, following recommendations from prior empirical evaluations of academic search systems \cite{zhang2011identifying}.
The search covers the {\em time period} until March 2026. This allowed inclusion of studies from the emergence of agent-based systems, \BlueText{with March 2026 used as the search end date to capture the most recent studies available during review execution.}
The search string below combines three concept groups taken from the research questions: agentic AI, the computing continuum, and architecture or deployment.\\

\fbox{%
\parbox{0.95\linewidth}{(“LLM-based agent” OR “LLM agent” OR “agentic AI”)
AND
(“edge” OR “fog” OR “cloud” OR “edge-cloud” OR “cloud-edge”)
AND
(“architecture” OR “deployment” OR “orchestration”)
}}\\

\BlueText{The search string selects papers that describe their own systems as ``LLM-based agents,'' ``LLM agents,'' or ``agentic AI.'' This improves precision, but relevant studies may use other labels, such as distributed LLM systems, federated LLMs, collaborative inference, or model placement, without using agent vocabulary. The search was therefore run on full text rather than only on titles and abstracts, and the snowballing stage was designed as a safeguard against such terminology differences, not as an optional extra step.}

\BlueText{We also checked directly how much the wording of the search string matters. EdgeShard \cite{edgeshard}, Galaxy \cite{galaxy}, and PerLLM \cite{perllm} are well-known systems that distribute LLM inference across devices, edge, and cloud without describing themselves as agent systems. All three are cited by studies in our corpus (\cite{P4}, \cite{P33}, and \cite{P30}, respectively), so backward snowballing reaches them. They are not primary studies here because they optimize the serving of a single model rather than a system of cooperating components (criterion EC2 in Table \ref{inc}), or are not peer reviewed (EC4). The full-text exclusion log in the replication package records 31 such decisions with the criterion applied and a short reason. This check reduces, but cannot remove, the risk that some multi-component systems described without agent vocabulary were missed; Section \ref{sec:threats} returns to this point.}

\begin{figure}
    \centering
    \includegraphics[width=0.5\linewidth]{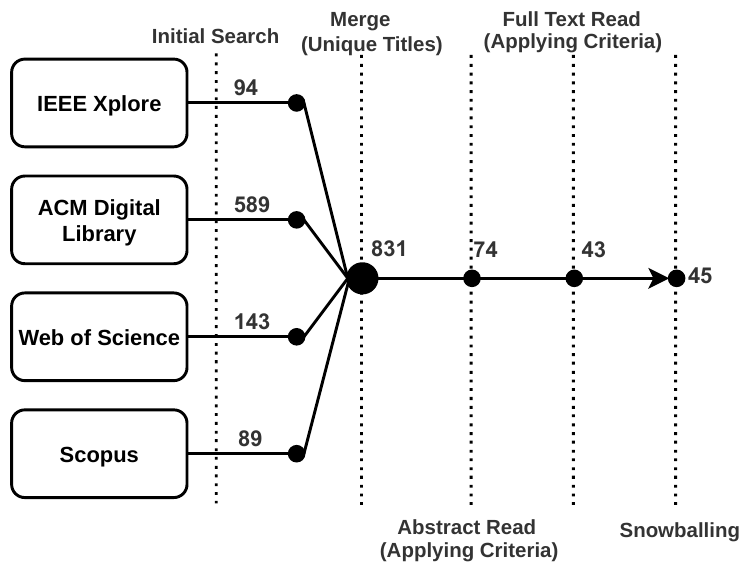}
    \caption{Search and selection process.}
    \label{search}
\end{figure}

The {\bf selection} process shown in Figure \ref{search} follows a staged filtering approach aligned with PRISMA \cite{page2021prisma}:
{\em i) Identification:} the four databases returned 915 records (IEEE Xplore 94, ACM Digital Library 589, Scopus 143, Web of Science 89); merging and removing duplicates left 831. The per-database yields differ mainly because each engine exposes different searchable fields; the exact strings and fields per database are listed in the replication package.
{\em ii) Screening:} apply inclusion/exclusion criteria (see Table \ref{inc}) to titles and abstracts. A paper was included only if it met all inclusion criteria and none of the exclusion criteria. The papers were divided among the authors, and each researcher's included papers were checked by the others to minimize bias. This step led to 74 papers.
{\em iii) Eligibility:} full-text assessment of the remaining studies, resulting in 43 papers. The 31 exclusions at this stage are logged with the criterion applied and a short reason.
{\em iv) Expansion:} apply backward and forward snowballing, which added two records and resulted in the 45 primary studies listed in Appendix~\ref{app:primary-studies} (Table~\ref{tab:primary-studies}).
\BlueText{The two snowballed records are the fine-tuning method \cite{P44} and the secondary study \cite{P45}. Both fail a strict reading of the exclusion criteria (EC2 and EC4, Table \ref{inc}), and we report their retention as a protocol deviation. We kept them because the three system papers of the federated family build directly on them, and they are marked as supporting records in Appendix \ref{app:consolidated-mapping}. Excluding them changes no family, no pattern, and none of the reported gaps, and every corpus-level count they enter is marked in the text.}
Two external experts in AI/LLM systems and edge computing reviewed the search and selection process. The consequences of the corpus size for the strength of our conclusions are discussed in Section \ref{sec:threats}.

We created an extraction form, and for each selected study, we {\em extracted} structured information on both architectural and operational aspects.
The form covers the system context and application domain, the agent roles and their coordination, model types and sizes, hardware, placement of components across device, edge, and cloud, adaptation and state management, governance and oversight, and the evaluation design. Using one form for all studies keeps the comparison consistent.

\begin{table}[t]
\scriptsize
\centering
\caption{Inclusion and exclusion criteria.}
\begin{tabular}{p{3cm} p{10cm}}
\toprule
\textbf{Inclusion Criteria} & \\
\midrule
Subject (IC1) & Multi-agent or multi-component LLM systems enabling mechanisms for cloud-edge-IoT deployment \\
Contribution (IC2) & Proposes, evaluates, or characterizes system architectures, deployment mechanisms, or design approaches relevant to multi-agent cloud-edge-IoT settings \\
Scope (IC3) & Addresses system-level, architectural, or deployment-level aspects \\
Type (IC4) & Peer-reviewed journals, conferences, workshops papers, or book chapters \\
Language (IC5) & English \\
Accessibility (IC6) & Full text available \\
\midrule
\textbf{Exclusion Criteria} & \\
\midrule
Subject (EC1) & Works not related to multi-agent systems in cloud-edge-IoT deployment contexts \\
Contribution (EC2) & Purely model-centric contributions without architectural or system-level implications for deployment \\
Scope (EC3) & No consideration of system design, orchestration, or deployment aspects \\
Type (EC4) & Non-peer-reviewed or secondary \\
Language (EC5) & Non-English publications \\
Accessibility (EC6) & Full text not accessible \\
\bottomrule
\end{tabular}
\label{inc}
\end{table}

The {\bf analysis} of the extracted data used two coding steps:
{\em i) Deductive coding:} mapping extracted data to predefined dimensions (architecture, deployment, application domain, evaluation practice).
{\em ii) Inductive coding:} identifying recurring mechanisms that the predefined dimensions did not capture, such as role-specialized collaboration, planner-worker orchestration, tool use, task decomposition, and governance-gated control loops.
The results are organized into three classifications, presented in Sections \ref{sec:rq1} to \ref{sec:rq3}: architecture families, deployment patterns, and evaluation practices.

\BlueText{Many systems fit more than one category, which makes classification partly a matter of judgment. We handled this with two rules. On the architecture dimension, each study was assigned to the one family that matches its main technical contribution. A second author checked every assignment, and disagreements were discussed until resolved. On the deployment and evaluation dimensions, we kept multiple codes instead of forcing one: five studies carry two deployment codes, and 28 carry two or more evaluation codes. Appendix \ref{app:consolidated-mapping} lists the complete coding for every study, so readers can check individual decisions.}

\BlueText{We did not pool quantitative results across studies. The primary studies use different tasks, datasets, hardware, network conditions, and metrics, so a meta-analysis of their reported numbers would not be meaningful \cite{kitchenham2015evidence}. The analysis instead reports how often each architecture, deployment mechanism, domain, and evaluation practice occurs, which combinations occur together, and which are missing. We make no claims that one architecture performs better than another.}

\BlueText{In addition to the classification, we recorded for each study how its system was evaluated (conceptual argument, simulation or dataset experiments, physical prototype or testbed, or production data), whether it compares against at least one baseline, whether it reports ablation experiments, whether code or data are public, and whether limitations are discussed. These records are reported in aggregate in Section \ref{sec:rq3} and per study in Appendix \ref{app:consolidated-mapping}.}

\BlueText{To strengthen procedural validity, the review protocol, extraction form, and coding structure were reviewed by the same two external experts. They were not additional screeners for inclusion decisions; they commented on the search strategy, the coding structure, and the interpretability of the results.}

\BlueText{Before full screening and extraction, we conducted a {\em pilot calibration} round to align the review team on borderline cases, especially papers that combined LLM reasoning with structured external components, edge deployment, or tool-mediated orchestration but did not always present themselves in uniform terminology. 
A subset of records was independently assessed by multiple members of the research team and disagreements were discussed until consensus was reached. The pilot was used to refine the inclusion and exclusion criteria, to stabilize the extraction form, and to agree on the meaning of concepts such as multi-agent organization, edge-first deployment, and governance-gated control. This reduced ambiguity before the full review.}

\BlueText{The resulting categories are not unique to this study. The planner-worker family matches the orchestrator structures described in surveys of LLM-based multi-agent systems \cite{S1,single}; the hybrid local/cloud family matches cloud-edge collaborative inference in edge-LLM surveys \cite{zheng2025edgellm,qu2025mei}; and the federated family follows the design space described in \cite{P45}. This correspondence with independently developed categories, the expert review of the protocol, and the dual checking of assignments are our main safeguards for the stability of the classification. No external benchmark exists against which the classification could be validated formally; Section \ref{sec:rq4} argues that building one is a priority for the field.}

\section{Architecture Families (RQ1)}
\label{sec:rq1}

Five architectural families recur in the primary studies (Table \ref{tab:arch-taxonomy}). The families overlap in practice, so each study is assigned to the family that matches its main contribution, following the rule described in Section \ref{sec:method}. The largest family does not build agent teams at all. It builds the supporting infrastructure, such as offloading, communication, memory, scheduling, and serving, that agent teams need in order to run on the continuum.

{\bf Edge-resident role-specialized teams.}
The studies in this family differ in how deep the specialization goes. At one end, specialization follows the pipeline. \cite{P3} chains a lightweight activity recognizer with hourly and daily LLM summarizers on local Raspberry Pi hardware. At the other end, \cite{P15} pushes specialization into the models themselves and assigns planning, execution, and interaction to separate expert SLMs under a finite-state controller for industrial IoT. Other studies split roles along physical concerns. \cite{P19} uses three agents for electric-vehicle threat analysis, charge estimation, and health anomalies, and \cite{P43} invokes LiDAR only when the camera-based interpretation is uncertain. Further studies apply the same structure to consensus, observability, and online safety \cite{P1,P34,P42}.

These systems are placed close to the data so that role boundaries can follow privacy zones, sensor layout, and the latency budget of the local decision. The drawback is that roles are fixed at design time; none of these systems reorganizes its team in response to load or context.

{\bf Hierarchical planner-worker orchestration.} 
\BlueText{The system in \cite{P2} is a concrete example. A chief agent interprets a natural-language telecom Operations Support Systems (OSS) reconfiguration request, translates it into a dependency-aware execution graph, and preserves ordering constraints such as allocating an IP address before tearing down a VLAN. Bounded atomic actions are then delegated to domain workers for service management, IP management, network control, and cloud orchestration. The design fits the continuum because cross-domain planning sits at the coordination layer, while narrowly scoped execution stays close to the domain APIs and the operational state that each worker controls.}

The same structure appears across domains. \cite{P7} organizes bounded agentic subsystems across end-to-end and specialized network domains in the AGILE-6G framework. \cite{P10} distributes roles across mobile devices and tiered edge servers and uses role-affinity scheduling with pipeline-parallel execution. \cite{P12} places a persona-based multi-tool architecture into the Radio Access Network (RAN) Intelligent Controller, combining traffic prediction and anomaly anticipation with safety-aligned rewards. \cite{P13} puts an LLM planner over perception, preprocessing, analytics, and optimization layers in smart manufacturing. \cite{P16} predicts service intents and injects them as global prompts into service-function-chain orchestration. \cite{P21} uses repair bots that monitor enterprise data pipelines and trigger remediation scripts. \cite{P29} partitions pathfinding problems into regions and assigns them to region-level LLM agents with reflection and replanning. \cite{P38} adds a policy-governed orchestration layer over sensor, Manufacturing Execution Systems (MES), and Enterprise Resource Planning (ERP) streams, with digital-twin validation and human review.

Across the family, strategic planning stays in high-level planner agents, and execution is delegated to narrowly scoped workers, tools, or resource tiers. This division maps well onto cloud-scale planning combined with edge-local execution.

\begin{table}
\centering
\caption{Dominant architectural families in mapping studies.}
\label{tab:arch-taxonomy}
\scriptsize
\setlength{\tabcolsep}{2pt}
\renewcommand{\arraystretch}{1.05}
\begin{tabularx}{\columnwidth}{p{0.21\columnwidth} p{0.40\columnwidth} p{0.34\columnwidth}}
\toprule
Architectural family & Defining idea & Primary studies \\
\midrule
Edge-resident role-specialized teams &
Multiple specialized agents operate close to data sources; privacy and immediacy dominate &
P1, P3, P15, P19, P34, P42, P43 \\\\
Hierarchical planner-worker orchestration &
A chief/planner/orchestrator decomposes goals and delegates to role-specific workers or tools &
P2, P7, P10, P12, P13, P16, P21, P29, P38 \\\\
Hybrid local-small / cloud-large collaboration &
Compact local models or gateways handle immediate tasks; cloud LLMs provide heavyweight reasoning or fallback &
P5, P8, P20, P23, P25, P26, P30, P31, P39, P41 \\\\
Federated and collaborative adaptation &
Learning or adaptation is distributed across edge nodes or nearby devices rather than centralized &
P14, P17, P33, P44, P45 \\\\
Substrate-oriented agentic control &
The main contribution is offloading, memory, communication, governance, serving, or migration rather than task semantics &
P4, P6, P9, P11, P18, P22, P24, P27, P28, P32, P35, P36, P37, P40 \\
\bottomrule
\end{tabularx}
\end{table}

{\bf Hybrid local-small and cloud-large collaboration.}
\cite{P5} uses a cloud 72B model to fine-tune a lightweight edge SLM, which then produces a one-round interaction graph that costs less communication than fully connected multi-agent exchanges. \cite{P8} monitors conditions with edge sensors and edge agents, performs context-aware reasoning with a cloud LLM, and retrieves temperament-aligned knowledge with a QA agent. \cite{P20} orchestrates specialized wireless-engineering tools instead of embedding that expertise in large models, which is what makes the system deployable at the edge. \cite{P23} places a selection policy at the edge that decides which model should answer a request based on time criticality and task complexity. \cite{P25} stays on constrained hardware and lets small edge models approach cloud-model reasoning quality through efficient test-time scaling. \cite{P26} decomposes tasks into subtasks, builds dependency graphs, and allocates the subtasks between a local SLM and a cloud LLM. \cite{P30} offloads between local and cloud models across modalities, tasks, and dialogues. \cite{P31} anonymizes Protected Health Information (PHI) and retrieves reports on the device, then uses cloud Retrieval-Augmented Generation (RAG) for the final advice. \cite{P39} deploys tiny models at the edge and large models in the cloud for offshore surveillance. \cite{P41} keeps an edge vector cache and learns whether to reuse cached answers or call the cloud LLM.

In this family, neither a purely local nor a purely cloud deployment is treated as sufficient. The systems combine small local models with heavier remote reasoning and define explicit rules for when the handoff happens.

{\bf Federated and collaborative adaptation.}
In this family, model adaptation itself is distributed rather than performed in the cloud. \cite{P14} federates LLM reporting across roadside inspection nodes with parameter-efficient fine-tuning and secure aggregation. \cite{P17} runs federated agent workloads on a heterogeneous K3s edge cluster and studies quantization and network-aware scheduling. \cite{P33} lets nearby personal devices fine-tune a model together through parallel adapters and activation caching. Two further studies entered through snowballing and play a supporting role. \cite{P44} contributes the memory-efficient fine-tuning method that edge-side adaptation builds on, and \cite{P45} maps the design space of federated and edge learning for LLMs. The three system papers depend directly on the techniques these two describe.

The family is small, but it is the only one in which adaptation after deployment is part of the architecture rather than a cloud-side activity.

{\bf Substrate-oriented agentic control.}
\cite{P4} optimizes task offloading and resource allocation for collaborative inference. \cite{P6} embeds a local tool-augmented LLM in a green network control loop. \cite{P9} studies the economics and security of outsourcing subtasks to third-party LLM agents. \cite{P11} contributes a knowledge base rather than a team structure: an on-device ontology and knowledge graph ground small-model intent reasoning, which makes reliable smart-home understanding possible without a cloud connection. \cite{P18} packages anomaly detection and energy forecasting as tools behind an LLM agent on an edge analytics platform. \cite{P22} analyzes the security of agent communication protocols. \cite{P24} makes edge agents behind NAT and firewalls reachable through shadowing and agent-to-agent communication. \cite{P27} turns privacy policies into concrete protective actions at the edge. \cite{P28} converts natural-language intents into access-control policies and device configuration in transient IoT. \cite{P32} optimizes cloud serving with multi-timescale routing and forecast-aware autoscaling. \cite{P35} evaluates a ReAct-style scheduler with scratchpad memory and constraint enforcement. \cite{P36} manages distributed agent memory indexed by intent. \cite{P37} combines the Model Context Protocol (MCP) with task-oriented networking to lower communication latency. \cite{P40} optimizes vehicular offloading and resource allocation with multi-agent deep reinforcement learning.

The other four families depend on this infrastructure work. These studies answer how edge agents communicate, where they store state, how they migrate and find one another, how they are protected, and how serving reacts to load.

\begin{figure}
    \centering
    \includegraphics[width=1\linewidth]{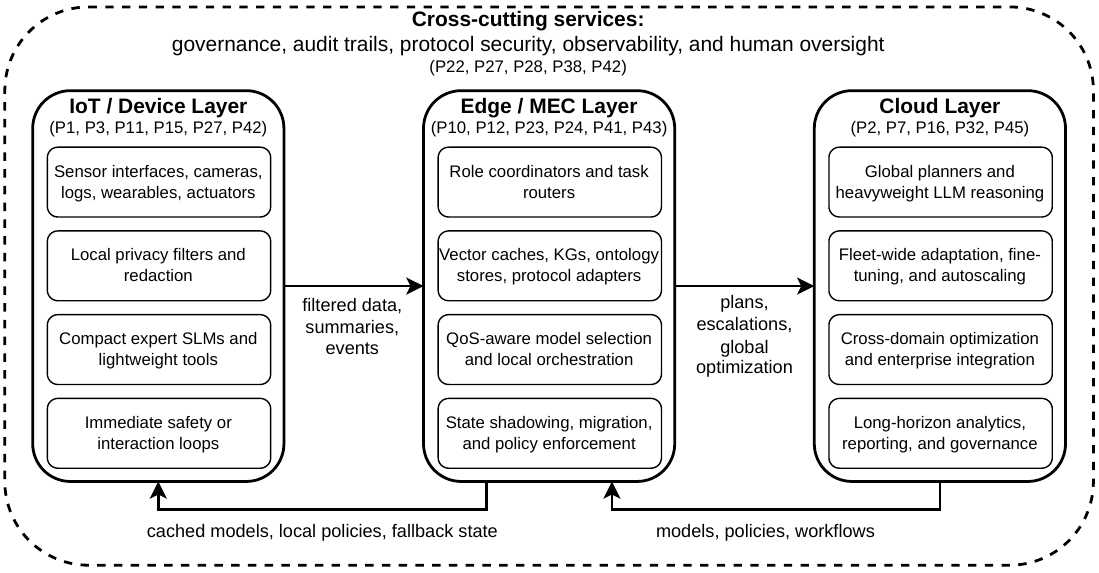}
    \caption{Reference model derived from the primary studies. The cloud-edge-IoT stack is both a deployment and a role hierarchy: privacy, latency, reasoning depth, and governance sit at different layers. \BlueText{Representative primary studies are shown inside the functional blocks they support, so the abstraction stays traceable to the studies.}}
    \label{fig:reference-architecture}
\end{figure}

Figure~\ref{fig:reference-architecture} combines the architectural observations into a three-layer reference model. The IoT/device layer hosts sensing, actuation, privacy filtering, and compact task-specific models for immediate interaction under latency and connectivity constraints \cite{muccini2018iot}. The edge/Multi-access Edge Computing (MEC) layer is the coordination layer. It hosts routers, vector or knowledge-graph caches, protocol adapters, and local orchestration services that sit between local autonomy and cloud escalation. The cloud layer holds heavyweight planning, long-horizon reasoning, cross-domain optimization, and lifecycle tasks such as fleet-wide fine-tuning and autoscaling. Upward arrows carry filtered observations and summaries toward plans and global optimization; downward arrows carry policies, models, and workflows back to the edge and device layers. Governance, audit, observability, and protocol security span all layers because they regulate the whole control path rather than a single tier.

\BlueText{
Architecture and deployment are separate questions. The same study can organize responsibilities one way and place them at runtime another way. A planner-worker system may still execute mostly locally, and a hybrid local/cloud system may still depend on governance gates or state shadowing. The next section therefore codes the primary studies again, by deployment mechanism rather than by architecture, so that organization and placement can be compared separately.
}

\paragraph{Strength of support by family.}
The families differ in how their claims are backed. All seven edge-resident systems are evaluated on physical devices or testbeds, which fits a family built on the claim that closeness to data matters. Eight of the ten hybrid local/cloud systems are evaluated on hardware as well. This family also reports several outcome types together most often, because the escalation decision forces quality, latency, and cost to be measured jointly \cite{P23,P26,P30,P41}. The planner-worker family has the weakest support relative to its ambition. Of its nine studies, four are evaluated only in simulation or on datasets and one is conceptual, so cross-domain orchestration, the central promise of this family, has the least hardware-based support. The federated family is small. Three of its five studies run on physical edge clusters, but among its three system papers only \cite{P17} releases artifacts, which matters for a family whose contribution is operational machinery. The infrastructure family spans the widest range, from production serving traces \cite{P32} to purely simulated offloading models \cite{P4,P40}, and results from these two ends should not be read as equally strong. Two observations cut across all families. Several systems name privacy as the reason for local placement \cite{P1,P3,P11,P15}, but direct privacy measurements, such as leakage, disclosure, or compliance rates, appear only in \cite{P27} and \cite{P31}; the other outcomes coded under the privacy, safety, and trust family concern safety or trust rather than privacy itself. \cite{P27} and \cite{P31}, in turn, do not report the latency and cost figures needed to judge what the protection costs. And no mapped study reorganizes its agent team at runtime, so nothing is known about how these designs behave when the team itself must change.

\section{Deployment Patterns (RQ2)}
\label{sec:rq2}

The architectural families above rely on a small set of recurring deployment patterns. Placement decisions concern more than where models run. Most systems also decide at which granularity work is routed: whole requests, subtasks, dialogue turns, memory accesses, or trainable parameters.
Whereas Table~\ref{tab:arch-taxonomy} classifies studies by architectural organization,
\BlueText{Table~\ref{tab:deployment-taxonomy} classifies them by deployment mechanism. Some studies appear under more than one pattern, for example when they combine local execution with governance gates, or hierarchical orchestration with federated adaptation.}
Two studies do not fix where components run and do not contribute a placement mechanism: the observability system of \cite{P34} and the scheduler evaluation of \cite{P35}. They are coded on the architecture, domain, and evaluation dimensions only, and their deployment cell in Table~\ref{tab:appendix-consolidated-mapping} carries a dash.

\begin{table}[htbp]
\centering
\caption{Deployment pattern taxonomy linking mechanisms to quality attributes.}
\label{tab:deployment-taxonomy}
\scriptsize
\begin{tabular}{p{0.20\linewidth}p{0.26\linewidth}p{0.20\linewidth}p{0.24\linewidth}}
\toprule
\textbf{Deployment pattern} & \textbf{Main mechanism} & \textbf{Main quality attributes} & \textbf{Primary studies}\\
\hline
Local-only execution and local actuation &
Reasoning and tool invocation stay on device or edge node &
Privacy, latency, offline continuity &
\BlueText{P1, P3, P11, P15, P18, P27, P42}\\\\
Edge-first with selective cloud escalation &
Complexity, confidence, or modality triggers remote reasoning &
Latency-accuracy trade-off, cost control, privacy preservation &
\BlueText{P8, P20, P23, P25, P26, P30, P31, P39, P41, P43}\\\\
Hierarchical edge/MEC/cloud orchestration &
Roles and subtasks are placed across tiers with schedulers or orchestrators &
Scalability, utilization, task completion, responsiveness &
\BlueText{P2, P4, P5, P6, P7, P10, P12, P13, P16, P17, P19, P21, P29, P32, P37, P40}\\\\
Federated or collaborative edge adaptation &
Learning updates or fine-tuning are distributed across clients or nearby nodes &
Personalization, bandwidth savings, privacy &
\BlueText{P14, P17, P33, P44, P45}\\\\
State continuity, shadowing, and migration &
Memory state, reachability, and agent images are externalized or migrated &
Resilience, mobility QoS, session continuity &
\BlueText{P22, P24, P36}\\\\
Governance-gated closed loops &
Policy, privacy, authorization, safety, or risk checks gate execution and sharing &
Trustworthiness, accountability, compliance &
\BlueText{P9, P27, P28, P31, P38, P42, P43}\\
\bottomrule
\end{tabular}
\end{table}

{\bf Local-only execution and local actuation.}
Seven studies keep reasoning and actuation on the device or the local edge node \cite{P1,P3,P11,P15,P18,P27,P42}. \cite{P11} avoids cloud dependence for smart-home intent understanding by combining an on-device SLM with an ontology and knowledge graph. \cite{P27} keeps privacy mediation at the edge and applies privacy-enhancing technologies (PETs), such as blurring and redaction, before any data is shared. \cite{P18} packages local analytical tools behind an LLM agent, partly so that small manufacturers can use the system without a cloud contract. In these systems, the local node interprets the data and makes the decisions. One caution applies to the pattern. Its studies treat keeping data local as the privacy mechanism, but within the pattern only \cite{P27} reports a directly measured privacy outcome; the other results coded under the privacy, safety, and trust family concern safety or trust rather than leakage or disclosure. The corpus-wide picture is the same: the only further study with a measured privacy outcome is \cite{P31}.

{\bf Edge-first with selective cloud escalation.}
\cite{P23} escalates at the request level. It predicts time criticality and task complexity and chooses an LLM accordingly. \cite{P41} makes a related decision between reusing an edge vector cache and calling the cloud. \cite{P26} and \cite{P30} escalate at the subtask level rather than for whole requests. \cite{P31} escalates only after protected health information has been transformed locally. \cite{P39} moves from tiny edge models to large cloud models when a scene requires more accurate reasoning, and \cite{P8} moves from edge observation to cloud reasoning and external knowledge retrieval. \cite{P20} escalates to specialized tools instead of a larger cloud model, which changes the cost structure without relying on model scale.
The recurring triggers are task complexity, model confidence, sensitivity of the raw data, and latency criticality.

{\bf Hierarchical edge/MEC/cloud orchestration.}
Sixteen studies place roles and subtasks across tiers with schedulers or orchestrators. In \cite{P2}, the planner decomposes intents and delegates across OSS domains. In \cite{P10}, role-affinity scheduling distributes roles across hierarchical edge resources. In \cite{P6}, a local tool-augmented LLM turns sustainability goals into MEC traffic-steering actions. In \cite{P16}, predicted service intents feed forward into service-function-chain deployment at the edge. In \cite{P32}, request routing and GPU autoscaling are optimized together over several timescales with traffic forecasting and integer linear programming, which shows that the cloud tier itself is adaptive rather than a static endpoint \cite{muccini2018self}. \cite{P37} improves the transport of task-related information between agents, so performance gains do not have to come from model placement alone. \BlueText{
The pattern also appears outside network management. \cite{P13} coordinates perception, preprocessing, analytics, and optimization layers through an LLM planner in smart manufacturing. \cite{P19} layers electric-vehicle functions so that edge computing supports low-latency inference by task-specific agents. \cite{P21} triggers repair agents over enterprise reporting workflows from a mobile-cloud orchestration layer. \cite{P29} separates global coordination from regional LLM agents for online pathfinding, and \cite{P40} formulates a cloud-edge-device framework for vehicular task offloading.
}

{\bf Federated and collaborative adaptation.} Three studies show variants of edge-side adaptation: federated LLM reporting in a roadside inspection pipeline \cite{P14}, federated LLM-agent workloads on a heterogeneous K3s cluster \cite{P17}, and collaborative fine-tuning of personal LLMs among nearby devices \cite{P33}. The fine-tuning method of \cite{P44} and the federated and edge learning design space of \cite{P45} supply the techniques that make these deployments practical.

{\bf State continuity, communication, and migration.} Only three studies fall under this pattern, although mobile and industrial IoT depend on it. \cite{P22} shows that the communication protocols themselves can be insecure in ways that change agent behavior. \cite{P24} makes agents reachable through shadowing instead of assuming that every agent is openly addressable. \cite{P36} treats agent memory as a distributed systems problem and lets agents negotiate memory operations through an intent-indexed framework.

{\bf Governance-gated closed loops.} Seven studies build governance into the deployment itself \cite{P9,P27,P28,P31,P38,P42,P43}. \cite{P28} converts high-level intent into fine-grained access-control policies and enforces them through proxies.
\BlueText{\cite{P9} conditions the outsourcing of agent tasks on cost-benefit analysis, skill verification, reputation, and security-risk assessment.}
\cite{P38} checks policy, risk, capability, and a digital twin before a manufacturing action is executed. The fusion controller of \cite{P43} has a similar effect: it decides whether the expected gain justifies spending bandwidth and latency on richer sensor fusion. This pattern matters most in regulated or high-risk IoT environments, which require a check of whether the system may act before it does.

Standards and risk frameworks outside the corpus point in the same direction. Edge computing was motivated from the start by latency, bandwidth, and privacy \cite{edgev}. The Model Context Protocol standardizes model-to-tool context exchange \cite{mcpanthropic}, the Agent2Agent (A2A) protocol targets inter-agent communication \cite{a2aprotocol}, and early security analyses of this protocol layer already document a growing attack surface \cite{hou2025mcp}. The NIST AI Risk Management Framework \cite{nist2023airmf} and the OWASP Top 10 for LLM applications \cite{owasp2025llm} both require explicit risk management, not only functional success.

Three weaknesses in the deployment literature are not visible from the taxonomy alone. First, every study implements its escalation trigger differently. \cite{P23} learns a time-criticality predictor, \cite{P26} decomposes dependency graphs, \cite{P30} uses modality- and dialogue-aware policies, and \cite{P41} uses cache-similarity thresholds. No study compares its trigger with another study's trigger, so it is unknown which escalation signal is most reliable. What is known is only that escalation beats purely local and purely cloud baselines within each study. Second, the local-only pattern is justified by privacy but rarely measured on privacy. Keeping data local is treated as protection in itself, which leaves leakage at inference time unexamined. Third, the state-continuity pattern rests on three studies, and none of them evaluates continuity under the conditions that motivate it, such as vehicular handover, intermittent uplinks, or NAT rebinding. The mechanisms that mobile and industrial IoT need most are currently the least tested.

\section{Evaluation Practices (RQ3)}
\label{sec:rq3}

{\bf Corpus profile.}
Before looking at what is measured, we summarize how the studies measure. Table~\ref{tab:evidence-profile} gives the profile of the 45 primary studies. The corpus is young. One study appeared in 2023 (the snowballed method paper \cite{P44}), one in 2024, thirty-one in 2025, and twelve in 2026, and journals and conferences are almost evenly represented (23 versus 22). The share of real deployments is substantial. Twenty-nine of the 45 studies (64\%) evaluate on physical devices, prototypes, or hardware testbeds, and two use production data \cite{P21,P32}. Twelve studies rely on simulation or dataset experiments only, and two are conceptual \cite{P7,P45}. Most studies compare against at least one baseline (42 of 45, 93\%), but only 15 report ablation experiments that isolate their own components. In most systems it therefore remains unclear how much the router, the planner, or the governance gate contributes on its own. Reproducibility is the weakest point. We found public code or data for six studies \cite{P17,P25,P26,P32,P35,P44}, one of them the snowballed method paper, or 13\% of the corpus, which is low for a systems-oriented literature and makes the comparison problem of Section~\ref{sec:rq4} worse. Finally, 24 of the 45 studies (53\%) address IoT, industrial, smart-home, healthcare, vehicular, or robotics applications. Nineteen of these are evaluated on physical IoT or edge hardware or on production data, and five are IoT-motivated but evaluated in simulation \cite{P13,P18,P29,P39,P40}. The remaining 21 studies contribute methods, security analyses, or infrastructure that are not tied to one application domain. In the rest of this paper, claims about cyber-physical deployment rest on the studies with physical IoT evaluations, and we state it where a conclusion rests on simulation alone.

The primary studies name their models and hardware. On the device and edge tiers, open models below 8B parameters dominate. Qwen, Llama, Phi, Gemma, and Mistral variants between 0.5B and 8B carry most of the local reasoning, with 7B and 8B the most common sizes, and several systems distill or fine-tune these models for their role \cite{P5,P15,P33}. On the cloud tier, GPT-4-class commercial models are the most frequent choice for heavy reasoning and appear in the evaluations of roughly half the corpus, followed by Gemini- and Claude-class models; open models above 70B appear in a few systems, such as the 72B teacher model of \cite{P5}. The papers name hardware just as often. Jetson boards appear in at least eight studies, smartphones in seven, Kubernetes or K3s edge clusters in six, Raspberry Pi devices in five, and datacenter GPUs in thirteen; a study counts toward several platforms when it uses several. These counts come from the experimental-setup sections of the papers. The pairing of small local models with large cloud models from Section~\ref{sec:rq1} is visible directly in this hardware split.

\begin{table}[t]
\centering
\caption{Profile of the 45 primary studies. Study-level values are listed in Appendix \ref{app:consolidated-mapping}.}
\label{tab:evidence-profile}
\scriptsize
\begin{tabular}{p{0.42\linewidth}p{0.14\linewidth}p{0.36\linewidth}}
\toprule
\textbf{Dimension} & \textbf{Count} & \textbf{Notes / studies}\\
\midrule
Publication year 2023 / 2024 / 2025 / 2026 & 1 / 1 / 31 / 12 & Search ends March 2026; 2023 is the snowballed P44\\
Journal / conference or workshop & 23 / 22 & Near-even venue split\\
Evaluated on production or field data & 2 & P21, P32\\
Evaluated on physical prototype or testbed & 29 & 64\% of corpus\\
Evaluated in simulation or on datasets only & 12 & P4, P9, P12, P13, P16, P18, P20, P29, P35, P39, P40, P44\\
Conceptual, no experimental evaluation & 2 & P7, P45\\
Baseline comparison reported & 42 & 93\% of corpus\\
Ablation of own components & 15 & 33\% of corpus\\
Public code or data & 6 & P17, P25, P26, P32, P35, P44 (13\%)\\
Limitations explicitly discussed & 32 & 71\% of corpus\\
IoT domain, physical or production evaluation & 19 & Basis for deployment claims\\
IoT domain, simulation only & 5 & P13, P18, P29, P39, P40\\
Domain-independent methods and infrastructure & 21 & Network orchestration and methods\\
\bottomrule
\end{tabular}
\end{table}

\vspace{1.5em}

Table \ref{tab:domain-overview} groups the primary studies by application domain. Two points stand out. First, 24 of the 45 studies address IoT, industrial, smart-home, healthcare, vehicular, or robotics settings.

Second, studies build their own domain-specific evaluations instead of reusing shared benchmarks. General agent benchmarks such as AgentBench exist \cite{bench}, but the primary studies rarely use them, and these benchmarks do not model continuum conditions such as network partitions, migration cost, or privacy-preserving preprocessing.

\begin{table}[t]
\centering
\caption{Application Domains.}
\label{tab:domain-overview}
\scriptsize
\setlength{\tabcolsep}{2pt}
\renewcommand{\arraystretch}{1.05}
\begin{tabularx}{\columnwidth}{p{0.35\columnwidth} p{0.25\columnwidth} p{0.35\columnwidth}}
\toprule
Domain & Studies & Emphasis \\
\midrule
Network and service orchestration &
P2, P6, P7, P10, P12, P16, P20, P23, P32, P41 &
Intent handling, routing, Quality of Service (QoS), autoscaling, 5G/6G and cloud-edge control \\
IoT, smart home, healthcare, and personal assistance &
P1, P3, P8, P11, P27, P28, P30, P31, P34, P37, P42 &
Local privacy, edge sensing, human-in-the-loop assistance, context-aware reasoning \\
Industrial/IIoT/manufacturing and enterprise automation &
P13, P14, P15, P18, P21, P33, P38 &
IIoT autonomy, maintenance, inspection, enterprise workflow, edge fine-tuning \\
Vehicular, robotics, and mobile cyber-physical systems &
P19, P24, P29, P39, P40, P43 &
Mobility, edge collaboration, low-latency control, multimodal sensing, migration \\
Cross-cutting methods, security, and enabling substrates &
P4, P5, P9, P17, P22, P25, P26, P35, P36, P44, P45 &
Offloading, communication, memory, quantized fine-tuning, federated/edge learning \\
\bottomrule
\end{tabularx}
\end{table}

\begin{table}[htbp]
\centering
\caption{Evaluation families observed in primary studies.}
\label{tab:evaluation-synthesis}
\scriptsize
\begin{tabular}{p{0.24\linewidth}p{0.36\linewidth}p{0.32\linewidth}}
\toprule
\textbf{Evaluation family} & \textbf{What is measured} & \textbf{Representative studies}\\
\hline
Latency, QoS, and responsiveness &
End-to-end latency, Time-To-First-Token, QoS scores, evacuation time, deployment delay, migration delay, response time, or turnaround time &
\BlueText{P2, P4, P6, P10, P12, P16, P17, P21, P23, P25, P26, P27, P28, P30, P32, P34, P35, P37, P40, P41, P43}\\\\

Task success, accuracy, or reasoning quality &
Recognition accuracy, success rate, recommendation quality, anomaly classification, path-planning quality, policy-generation accuracy, or reasoning accuracy &
\BlueText{P1, P2, P3, P5, P8, P11, P13, P15, P16, P18, P19, P20, P21, P26, P28, P29, P30, P34, P35, P38, P42, P43}\\\\

Privacy, safety, trust, and explainability &
PHI protection, governance compliance, outage prevention, interpretable decision support, protocol security, security-risk assessment, or explainable mitigation &
\BlueText{P1, P9, P11, P12, P14, P19, P22, P27, P31, P38, P42}\\\\

Cost, energy, bandwidth, and resource efficiency &
GPU-hours, bandwidth use, energy consumption, memory footprint, token consumption, communication load, API cost, or resource utilization &
\BlueText{P4, P5, P6, P9, P14, P17, P25, P26, P30, P32, P33, P35, P39, P40, P41, P43, P44, P45}\\\\

State continuity and systems resilience &
Memory-management overhead, session continuity, migration cost, communication robustness, reachability, or recovery behavior &
\BlueText{P22, P24, P36, P37, P41}\\
\bottomrule
\end{tabular}
\end{table}

Table~\ref{tab:evaluation-synthesis} groups the measured outcomes into five families. It also matters which outcome families are reported together. Latency and efficiency usually appear together, because both come out of the same offloading or scheduling experiment.
\BlueText{Task success is often reported without the latency, cost, privacy, or resilience context needed for comparison, although 28 studies report more than one outcome family, for example success together with latency or with cost.}
Privacy and governance results rarely come with standard latency or cost numbers, so it is hard to tell whether a privacy-preserving design pays for itself.
\BlueText{State continuity is the most isolated family; few studies report memory, migration, reachability, cache reuse, or robustness outcomes.}
This reporting pattern, more than any single result, is why papers in this area are hard to compare.

The reported numbers show real gains, each measured against its own baseline. Efficiency-oriented studies report latency reductions in the tens of percent \cite{P4,P17,P25,P32}. The federated deployments quantify their savings against study-specific baselines: \cite{P17} reports 75\% less network bandwidth, 45\% lower end-to-end latency, and 61\% less energy per inference than a naive deployment, and \cite{P14} reports 84\% less communication overhead than full model synchronization. At cloud scale, \cite{P32} reports up to 25\% GPU-hour savings and an 80\% reduction in auto-scaling waste against a production baseline, or about \$2.5M per month. Task-quality results cluster between 80\% and 95\% across recognition, retrieval, smart-home reasoning, and industrial scenarios \cite{P3,P8,P11,P15}, with larger relative gains when domain knowledge is moved into tools \cite{P20} or when corner cases are evaluated separately \cite{P43}. These numbers cannot be compared across papers. Baselines, hardware, and workloads differ, and few papers report enough of the joint efficiency, quality, and privacy profile for a reader to reconstruct the trade-off.

The privacy, safety, and governance results are fewer but methodologically clear. \cite{P12} evaluates safety in realistic 5G scenarios and reports zero outages under high load, against 8.4\% for fixed-power baselines and 3.3\% for a competing agent baseline. \cite{P27} evaluates policy-aware PET selection on real traces and reports near-perfect compliance at about three seconds of end-to-end latency. \cite{P28} reports fully accurate policy generation in its transient-IoT configuration tasks. \cite{P31} separates PHI anonymization from cloud reasoning, so the privacy guarantee depends on the deployment rather than on the model. \cite{P22} shifts the evaluation target from model behavior to protocol security and exposes weaknesses in authentication, authorization, integrity, confidentiality, and availability, which matter more as deployments adopt open protocols such as MCP.

\BlueText{Individual studies show which outcomes dominate in which part of the design space. \cite{P3} reports 95.8\% activity-recognition accuracy, with summary quality comparable to a cloud baseline, while all processing stays local. \cite{P11} reports 82.2\% task success in general smart-home scenarios and 76.6\% in complex ones through ontology-supported on-device reasoning. \cite{P17} reports large bandwidth, latency, and energy reductions for federated edge deployments. \cite{P23} improves QoS by selecting models by time criticality and complexity. \cite{P32} lowers GPU-hour use and auto-scaling waste at cloud scale. \cite{P41} improves QoS with a vector cache in front of the cloud, and \cite{P43} improves corner-case detection while sending less data. Each result is credible in its own setting. Because the tasks, infrastructures, and outcome sets differ, the results cannot be weighed against each other.
}

\BlueText{The coding also shows why 28 studies appear in more than one evaluation family. \cite{P2} reports execution time and error behavior together with successful intent-driven OSS execution. \cite{P5} evaluates reasoning accuracy and communication cost. \cite{P26} and \cite{P30} optimize quality, latency, and cost together in local-cloud settings. \cite{P43} combines detection quality with the latency and bandwidth effects of selective sensing. These evaluations are the most useful ones in the corpus because they expose trade-offs instead of isolated scores.}

These observations motivate the reporting checklist in Section \ref{sec:rq4}. The ingredients for comparison exist, since latency, success, efficiency, privacy, and resilience are each measured somewhere in the corpus, but they are reported in non-overlapping subsets under non-overlapping conditions. Closing the gap requires varying task difficulty and deployment conditions together, which is what Figure~\ref{fig:benchmark} describes.

\BlueText{Appendix Table~\ref{tab:appendix-consolidated-mapping} lists the coding of all 45 primary studies on the four dimensions, together with each study's evaluation setting and artifact availability. It also shows which studies span several deployment or evaluation categories.}

\section{Open Challenges (RQ4)}
\label{sec:rq4}
\begin{figure}
    \centering
    \includegraphics[width=\linewidth]{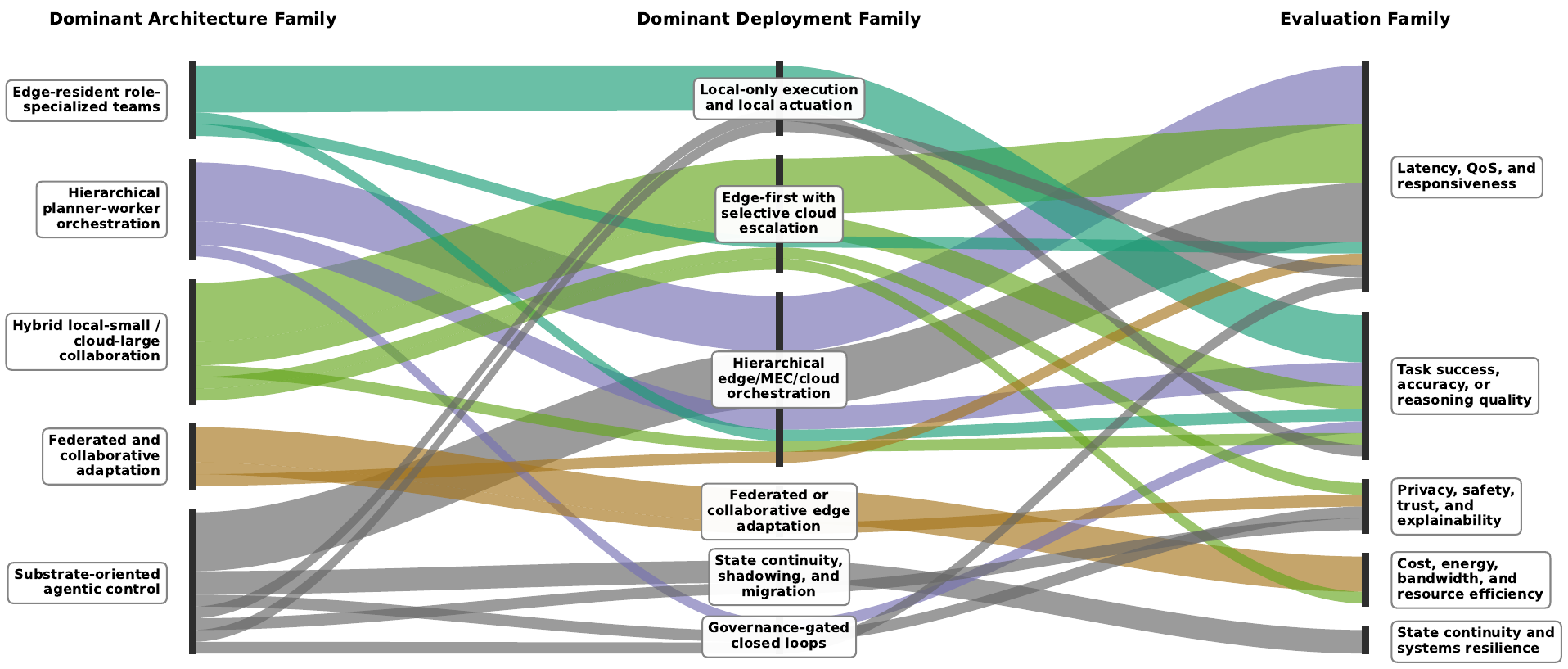}
    \caption{\BlueText{Flows from architecture family (left) through deployment pattern (middle) to evaluation family (right). Each primary study contributes one flow; for studies with several deployment or evaluation codes, the first-listed code of Appendix Table \ref{tab:appendix-consolidated-mapping} is drawn, and the complete assignment is in the appendix.}}
    \label{fig:overview}
\end{figure}

The open problems that emerge from the mapping concern the research practice of the field as much as its algorithms.
The first open problem is {\bf benchmark fragmentation}. Many individual papers are evaluated carefully, but almost every study builds its own evaluation environment. Five studies benchmark offloading, each in its own way \cite{P4,P16,P29,P39,P40}.
\BlueText{\cite{P4} and \cite{P40}, for example, both study offloading in edge or vehicular settings, and both report gains in delay and efficiency, but they use different environments, optimization goals, baselines, and outcome sets, so their results cannot be ranked against each other.} %
The same holds for local and hybrid assistance \cite{P1,P3,P8,P11,P31} and for industrial settings \cite{P13,P15,P18,P38}. The studies make progress individually, but later work cannot build on earlier numbers. A benchmark for this area would have to combine agent tasks with continuum variables such as bandwidth, round-trip time, memory synchronization policy, and local privacy transformations.

The second open problem is {\bf state continuity under mobility and disconnection}. Only three studies make reachability, memory consistency, or migration their main concern \cite{P22,P24,P36}, and \cite{P41} adds cache-reuse results inside the escalation pattern (see Figure \ref{fig:overview}).
These are the conditions that matter most in mobile, vehicular, and industrial deployments. Most architecture papers assume better connectivity than their target settings can offer.

The third open problem is {\bf protocol and tool security}. \cite{P22} is the only systematic security analysis of agent communication protocols in the corpus. \cite{P24} and \cite{P37} improve reachability and communication efficiency, and \cite{P27} and \cite{P28} add governance layers, but adversarial evaluation is rare. At the same time, MCP \cite{mcpanthropic} and A2A \cite{a2aprotocol} are being adopted quickly, security analyses of this protocol layer already list authentication, authorization, and context-poisoning weaknesses \cite{hou2025mcp}, and application-security guidance counts excessive agency and insecure tool use among the main LLM risks \cite{owasp2025llm}.
Edge-resident teams are absent from this line of work, although privacy is one of their main motivations (see Figure \ref{fig:overview}).

The fourth open problem is {\bf federated lifecycle management}. Three studies show that collaborative edge adaptation works \cite{P14,P17,P33}, and the methods they build on explain why it has become affordable \cite{P44,P45}.
What is missing is guidance on when to federate, how to handle heterogeneous edge hardware, how to evaluate update drift, and how to align federated fine-tuning with governance requirements.

The fifth open problem is {\bf human oversight and traceable decisions}. Several studies build human-in-the-loop assistance, explanation, or policy traceability into their architectures \cite{P8,P13,P19,P21,P27,P38}. There is no agreement yet on what should be explainable at the device, edge, or cloud level, or on how to audit decisions that cross layers.

\BlueText{The sixth open problem is {\bf artifact availability}. Six of the 45 studies publish code or data. Without artifacts, results cannot be re-measured when models, hardware, or network conditions change, and a shared benchmark cannot start from reference implementations. Venues in this area could treat artifact release, or a stated reason why sharing is impossible, as part of the review bar. The artifact evaluation of \cite{P25} shows that this is practical.}

\begin{figure}
    \centering
    \includegraphics[width=0.7\linewidth]{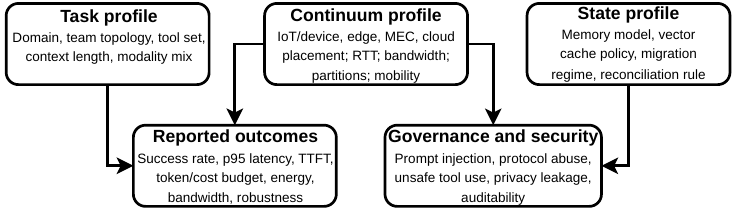}
    \caption{Suggested benchmark harness for cloud-edge-IoT multi-agent LLM systems.}
    \label{fig:benchmark}
\end{figure}

Figure \ref{fig:benchmark} summarizes what a benchmark for this area would have to control: a task profile, a continuum profile, a state profile, a governance and security profile, and a fixed set of reported outcomes.
\BlueText{We have not implemented this benchmark, and the mapping shows why building it exceeds a single group: the workloads, hardware, and governance settings are spread across domains, and only six studies release artifacts to build on. The reporting side, however, can be adopted now. Table \ref{tab:benchmark-operationalization} lists, for each profile, the variables a study should fix or report. A study that reports them can be compared with other such studies even before a shared harness exists. If a harness is built later, the same variables define its configuration space. A shared setup would also show whether a system is fast because of its design or simply because it offloads more, skips privacy steps, or faces an easier network.}

\begin{table}[t]
\caption{\BlueText{Reporting checklist derived from the mapping. For each profile of Figure~\ref{fig:benchmark}, the table lists the variables a study should fix or report so that its results can be compared with other studies.}}
\label{tab:benchmark-operationalization}
\centering
\scriptsize
\begin{tabular}{p{0.20\linewidth} p{0.40\linewidth} p{0.30\linewidth}}
\hline
\textbf{Benchmark profile} &
\textbf{Variables that should be fixed or reported} &
\textbf{Example values or ranges} \\
\toprule
Task profile &
Task type, tool complexity, modality, deadline class, interaction horizon &
Local intent reasoning; multimodal fusion; one-step vs.\ long-horizon tasks; soft vs.\ hard deadlines \\\\
Continuum profile &
Device class, edge class, cloud class, RTT, bandwidth, packet loss, intermittence &
Phone, Raspberry Pi, Jetson, edge server, cloud VM; 10--200\,ms RTT; intermittent uplink \\\\
State profile &
Memory type, cache type, shadow/twin state, migration frequency, recovery target &
Stateless; session memory; vector cache; device shadow; failover within X seconds \\\\
Governance and security profile &
Authentication mode, tool permissions, human-approval points, audit logging, data sensitivity &
OAuth/token authentication; allow-listed tools; human sign-off before actuation; audit logging enabled \\\\
Reported outcomes &
Latency, task quality, energy, bandwidth, runtime cost, resilience, privacy/safety indicators &
p50/p95 latency; success rate; Joules; MB transferred; cost per task; recovery success rate\\
\bottomrule
\end{tabular}
\end{table}

{\bf Design guidance for practitioners.}
The mapping also supports practical guidance. Table~\ref{tab:design-guidance} lists common IoT constraints, the deployment pattern the corpus associates with each, example studies, and the costs the studies report or concede. Three notes on use. The examples are taken from studies with physical or production evaluations wherever possible, so each row is backed by at least one real deployment. The rows can be combined; systems often pair an escalation policy with governance gates, and Appendix \ref{app:consolidated-mapping} shows which studies combine which patterns. The last column records the trade-offs the studies themselves measured or admitted.

\begin{table}[t]
\centering
\caption{Deployment guidance for common IoT constraints, based on the mapped studies. Example studies have physical or production evaluations where available.}
\label{tab:design-guidance}
\scriptsize
\begin{tabular}{p{0.24\linewidth}p{0.20\linewidth}p{0.14\linewidth}p{0.32\linewidth}}
\toprule
\textbf{Constraint} & \textbf{Suggested pattern} & \textbf{Examples} & \textbf{Reported costs and risks}\\
\midrule
Raw data must not leave the premises (health data, minors, household audio and video) & Local-only execution; add governance gates and local PETs where data must be shared & \cite{P11,P27,P42} & Local model quality ceiling; privacy still has to be measured; plan for local model updates. \cite{P31} shows the variant in which only anonymized data leaves\\\\
Sub-second interaction on constrained hardware & Local or edge-first execution with expert SLMs and tools & \cite{P3,P15,P23,P25} & Limited reasoning depth; needs a calibrated confidence signal to know when local is not enough\\\\
Bursty task complexity under a cost ceiling & Edge-first with request- or subtask-level escalation & \cite{P23,P26,P30,P41} & Escalation triggers differ per study and are not calibrated across systems; mis-routing hurts cost and quality\\\\
Goals that span several subsystems & Planner-worker orchestration & \cite{P2,P10,P38} & Planner is a single point of failure; verify worker outputs; hardware-based support still thin\\\\
Fleet personalization without centralizing data & Federated or collaborative adaptation & \cite{P14,P17,P33} & Heterogeneous edge hardware and update drift; align fine-tuning with governance requirements\\\\
Mobility, NAT, intermittent connectivity & State shadowing and externalized memory & \cite{P24,P36} & Weakest support of all rows; continuity under real disconnections is untested\\\\
Regulated or safety-critical actuation & Governance-gated control with human approval & \cite{P28,P38,P42,P43} & Gate latency and audit overhead; needs an explicit policy model and logging from the start\\
\bottomrule
\end{tabular}
\end{table}

\section{Threats to Validity}
\label{sec:threats}

\BlueText{{\bf Construct validity.} The main threat is the search strategy. The search string selects studies that call their systems LLM agents, LLM-based agents, or agentic AI, so relevant work under other labels, such as distributed LLM systems, federated LLMs, collaborative inference, or model placement, may be underrepresented. We reduced this risk in three ways. The search ran on full text in four databases, not only on titles and abstracts. Backward and forward snowballing was applied from the included studies. And we checked directly that three well-known systems from the adjacent terminology (EdgeShard, Galaxy, PerLLM) are reachable from the corpus by snowballing, as described in Section \ref{sec:method}; they are excluded by our criteria because they optimize a single model, not because the search missed their vocabulary. The string also omits the bare terms multi-agent and IoT, so coverage of such work depends on the papers using one of the agent terms somewhere in their full text. The remaining risk concerns multi-component systems that are described without any agent vocabulary. We cannot rule such cases out, and the corpus should not be read as complete in that direction.}

\BlueText{{\bf Internal validity.} Selection, extraction, and classification involve judgment. Several studies show traits of more than one architectural family or deployment pattern. We reduced the risk of bias with a predefined protocol, one shared extraction form, a pilot round on borderline cases, checking of every architecture assignment by a second author, and an external expert review of the search, extraction, and analysis steps. The record of how each study was evaluated relies on what the papers state, and borderline judgments, for example between a hardware testbed and a realistic emulation, remain possible. We did not compute an inter-rater agreement statistic; the complete coding is published, so others can re-code the corpus and measure agreement against it. Some subjectivity cannot be removed when heterogeneous systems are mapped into shared categories.}

\BlueText{{\bf Conclusion validity.} The taxonomy is one reasonable structure for these 45 studies, not the only one. Other researchers could emphasize other dimensions or classify borderline studies differently. The reported evaluations differ in workloads, datasets, hardware, network conditions, and metrics, so we do not compare performance across studies and we make no claims that one architecture outperforms another. Our conclusions are statements about frequencies, combinations, and gaps. Each family and pattern rests on at least three primary studies. Removing the two records added by snowballing changes no category and no reported trend, and the corpus-level counts they enter (publication years, artifact count, size of the federated family) are marked where they are reported.}

\BlueText{{\bf External validity.} Most primary studies appeared in 2025 and 2026, and 45 studies is a modest base for a fast-moving field. The families, patterns, and gaps reported here describe the literature up to March 2026 and should be read as a snapshot, not as a lasting characterization. Since 24 of the 45 studies are grounded in IoT, industrial, smart-home, healthcare, vehicular, robotics, or sensor-cloud settings, the findings may also not carry over to multi-agent LLM systems outside continuum settings.}

\BlueText{{\bf Reliability.} The protocol, search strings, selection log, extraction sheet, and classifications are archived with a persistent DOI, so the selection and coding can be checked and re-run. Mapping studies still involve interpretation, and different researchers might decide a small number of borderline cases differently. We expect the overall structure to be stable under such differences, and the sensitivity check above supports this for the snowballed studies.}

\section{Conclusion}
\label{sec:conclusion}

\BlueText{This paper reported a systematic mapping study of multi-agent LLM systems on the cloud-edge-IoT continuum. From 831 records retrieved up to March 2026, we selected 45 primary studies, classified each by architecture, deployment mechanism, application domain, and evaluation practice, and recorded how each study evaluates its system.}

\BlueText{The mapped studies repeat a small set of designs. Five architectural families recur, from specialized agent teams at the edge to the infrastructure that carries them, and deployment relies on six recurring mechanisms, from local-only execution to governance-gated control. More than half of the studies are set in IoT, industrial, healthcare, smart-home, vehicular, or robotics applications, and 29 of the 45 are evaluated on physical hardware, mostly with open models below 8B parameters on the device and edge tiers. These findings describe the literature up to March 2026.}

\BlueText{The main obstacle the mapping exposes is comparability. Studies report gains in latency, task quality, efficiency, privacy, or resilience, but under different workloads, hardware, network conditions, and metrics, and only six studies release artifacts. Privacy in particular motivates many designs but is directly measured in only two studies. Until studies report their deployment conditions in a shared form, new results will remain difficult to compare with old ones. The reporting checklist in Section \ref{sec:rq4} is a step that individual authors can take now, without waiting for a common benchmark.}

\BlueText{For IoT researchers and practitioners, the practical questions are how to divide responsibilities among devices, edge resources, and cloud services, and which governance mechanisms should regulate the result. The taxonomy, the study-level records, and the guidance in this paper are intended as a starting point for those decisions, and the archived protocol and data allow both to be checked and extended.}

\BlueText{
\section*{Data Availability}
The review protocol, search strings, study-selection log, coded data-extraction sheet, primary-study overview table, consolidated mapping table, and supplementary materials supporting the findings of this study are publicly available through the Zenodo replication archive. The Zenodo link is available at \url{https://doi.org/10.5281/zenodo.20631424}.
The archive contains the information required to reproduce the study selection, to verify the reported classifications, and to inspect the data behind the taxonomy and the analysis in this paper.
}

\section*{CRediT Authorship Contribution Statement}
\textbf{Mahyar Tourchi Moghaddam:} Conceptualization, Methodology,
Investigation, Data curation, Validation, Formal analysis,
Writing - original draft, Writing - review \& editing, Supervision,
Project administration.
\textbf{Tiziano Santilli:} Investigation, Data curation, Validation,
Writing - review \& editing.
\textbf{Leo Wilhelm Lierse:} Investigation, Data curation, Validation,
Writing - review \& editing.
\textbf{Sebastian Werner:} Investigation, Data curation, Validation,
Writing - review \& editing.
\textbf{Francesco Daghero:} Investigation, Data curation, Validation,
Writing - review \& editing.

\section*{Declaration of the Use of Generative AI}
During the preparation of this work, the authors used AI-based tools (Grammarly and Claude) for language editing of author-written text (grammar, shortening, rewording), formatting of LaTeX tables, and checks. The search, screening, inclusion and exclusion decisions, data extraction, coding, classification, and conclusions are the authors' own. Authors read the full text of every primary study, and every statement this paper makes about a primary study was checked against that study. 
All AI-edited text was reviewed and revised by the authors, who take full responsibility for the content of this article, following ARISMA guidelines \cite{moghaddam2026arisma}.

\section*{Declaration of Competing Interest}
The authors declare that they have no known competing financial interests or personal relationships that could have appeared to influence the work reported in this paper.

\renewcommand{\refname}{Primary Studies of the Mapping}

\appendix

\section{Overview of the Primary Studies}
\label{app:primary-studies}
\begin{scriptsize}
\setlength{\tabcolsep}{2.5pt}
\renewcommand{\arraystretch}{1.12}
\begin{longtable}{L{0.04\linewidth} L{0.1\linewidth} L{0.33\linewidth} L{0.20\linewidth} L{0.2\linewidth} L{0.03\linewidth}}
\caption{Overview of the 45 primary studies included in the systematic mapping study.}\label{tab:primary-studies}\\
\toprule
\textbf{ID} & \textbf{First author, year} & \textbf{Full paper title} & \textbf{Mapped application domain} & \textbf{Publication venue} & \textbf{Type}\\
\midrule
\endfirsthead
\caption[]{Overview of the 45 primary studies included in the systematic mapping study (continued).}\\
\toprule
\textbf{ID} & \textbf{First author, year} & \textbf{Full paper title} & \textbf{Mapped application domain} & \textbf{Publication venue} & \textbf{Type}\\
\midrule
\endhead
\midrule
\multicolumn{6}{r}{\emph{Continued on next page}}\\
\endfoot
\bottomrule
\endlastfoot
P1 & Fu (2025) & MALLM: Multi-Agent Decision-Making with LLMs for Multi-User Edge-Sensor Environments & IoT, smart home, healthcare, and personal assistance & ACM SIGMETRICS Performance Evaluation Review & J\\
P2 & Lee (2025) & Intent-Driven Autonomous Reconfiguration for Telecom OSS: A Generative-AI-Empowered Hierarchical Multi-Agent Architecture & Network and service orchestration & Asia-Pacific Network Operations and Management Symposium (APNOMS) & C\\
P3 & Inohara (2025) & Collaborative Lightweight LLM Agents for Daily Activity Summarization on Edge Devices & IoT, smart home, healthcare, and personal assistance & International Conference on Intelligent Environments (IE) & C\\
P4 & Yang (2025) & Task Offloading for Collaborative Inference of LLM Agents at the Network Edge & Cross-cutting methods, security, and enabling substrates & IEEE International Conference on Mobile Ad-Hoc and Smart Systems (MASS) & C\\
P5 & Zeng (2026) & Multi-Agent Collaborative Reasoning via Cloud-Edge Framework & Cross-cutting methods, security, and enabling substrates & IEEE Network & J\\
P6 & Moreira (2026) & AGORA: Agentic Green Orchestration Architecture for Beyond 5G Networks & Network and service orchestration & IEEE Access & J\\
P7 & Antonopoulos (2025) & AGILE-6G: Agentic AI for Autonomous Management of 6G Network/Application Services & Network and service orchestration & IEEE Network & J\\
P8 & Zhang (2025) & A Temperament-Aware IoT Parenting System with Multi-Agent LLM Reasoning and Personalized Knowledge Retrieval & IoT, smart home, healthcare, and personal assistance & IEEE World Forum on Internet of Things (WF-IoT) & C\\
P9 & Bhatt (2025) & COALESCE: Economic and Security Dynamics of Skill-Based Task Outsourcing Among Team of Autonomous LLM Agents & Cross-cutting methods, security, and enabling substrates & Cyber Awareness and Research Symposium (CARS) & C\\
P10 & Yu (2026) & CORE: Toward Ubiquitous 6G Intelligence Through Collaborative Orchestration of Large Language Model Agents Over Hierarchical Edge & Network and service orchestration & IEEE Communications Magazine & J\\
P11 & Jeong (2025) & On-Device Intent Reasoning for Smart Home Agents Via Ontology-Augmented sLLMs & IoT, smart home, healthcare, and personal assistance & IEEE Access & J\\
P12 & Salama (2025) & Edge Agentic AI Framework for Autonomous Network Optimisation in O-RAN & Network and service orchestration & IEEE International Symposium on Personal, Indoor and Mobile Radio Communications (PIMRC) & C\\
P13 & Farahani (2026) & Hybrid agentic AI and multi-agent systems in smart manufacturing & Industrial/IIoT/manufacturing and enterprise automation & Journal of Manufacturing Systems & J\\
P14 & Dell'Acqua (2025) & Agentic Federated LLMs at the Edge for Promptable Guardrail Inspection with YOLO-E & Industrial/IIoT/manufacturing and enterprise automation & IEEE/ACM International Conference on Utility and Cloud Computing & C\\
P15 & Yang (2026) & LEAF: A Lightweight Edge Agent Framework with Expert SLMs for the Industrial Internet of Things & Industrial/IIoT/manufacturing and enterprise automation & Computers, Materials, \& Continua & J\\
P16 & Sun (2026) & Generative Intent Prediction Agentic AI empowered Edge Service Function Chain Orchestration & Network and service orchestration & IEEE Transactions on Cognitive Communications and Networking & J\\
P17 & Tadi (2025) & Performance and Cost Optimization of Federated LLM Agents in Edge Computing Environments & Cross-cutting methods, security, and enabling substrates & International Conference on Foundation and Large Language Models (FLLM) & C\\
P18 & Chen (2025) & Voltage Anomaly Detection and Forecasting of Electrical Energy Consumption for Machine Tools using an AI-based Agent Framework & Industrial/IIoT/manufacturing and enterprise automation & IEEE Annual Congress on Artificial Intelligence of Things (AIoT) & C\\
P19 & Dif (2025) & Towards Trustworthy Agentic IoEV: AI Agents for Explainable Cyberthreat Mitigation and State Analytics & Vehicular, robotics, and mobile cyber-physical systems & IEEE Conference on Local Computer Networks (LCN) & C\\
P20 & Zhang (2026) & Small Models, Big Impact: Tool-Augmented AI Agents for Wireless Network Planning & Network and service orchestration & IEEE Communications Magazine & J\\
P21 & Patlolla (2026) & Intelligent Repair Bots for Power BI: An Agentic Automation Approach & Industrial/IIoT/manufacturing and enterprise automation & IEEE International Conference on AI in Cybersecurity (ICAIC) & C\\
P22 & Louck (2025) & Security Analysis of Agentic AI Communication Protocols: A Comparative Evaluation & Cross-cutting methods, security, and enabling substrates & ACM Transactions on AI Security and Privacy & J\\
P23 & Tuli (2025) & SELA: Smart Edge LLM Agent to Optimize Response Trade-offs of AI Assistants & Network and service orchestration & Proceedings of the ACM on Interactive, Mobile, Wearable and Ubiquitous Technologies & J\\
P24 & Sarkar (2025) & Agent Shadowing: Edge-Driven Seamless Integration for Edge and Cloud AI Agents & Vehicular, robotics, and mobile cyber-physical systems & International Workshop on Middleware for IT/OT Integration & C\\
P25 & Chen (2026) & FastTTS: Accelerating Test-Time Scaling for Edge LLM Reasoning & Cross-cutting methods, security, and enabling substrates & ACM International Conference on Architectural Support for Programming Languages and Operating Systems & C\\
P26 & Shao (2025) & Division-of-Thoughts: Harnessing Hybrid Language Model Synergy for Efficient On-Device Agents & Cross-cutting methods, security, and enabling substrates & ACM Web Conference 2025 & C\\
P27 & Khan (2026) & Governance at the Edge: Agent-Driven Privacy Mediation for Mobile and IoT Data & IoT, smart home, healthcare, and personal assistance & International Workshop on Mobile Computing Systems and Applications & C\\
P28 & Shastri (2025) & LLM-Driven Auto Configuration for Transient IoT Device Collaboration & IoT, smart home, healthcare, and personal assistance & ACM/IEEE Symposium on Edge Computing & C\\
P29 & Yang (2025) & DREAM: Distributed Regional Efficient Agent Management with LLMs for Online Multi-Agent Pathfinding & Vehicular, robotics, and mobile cyber-physical systems & ACM SIGOPS Operating Systems Review & J\\
P30 & Yuan (2025) & Local-Cloud Inference Offloading for LLMs in Multi-Modal, Multi-Task, Multi-Dialogue Settings & IoT, smart home, healthcare, and personal assistance & International Symposium on Theory, Algorithmic Foundations, and Protocol Design for Mobile Networks and Mobile Computing & C\\
P31 & Weerasekara (2025) & Privacy-Preserving Medical Advising System on Mobile Devices: On-Device PHI Anonymization, Medical Report Retrieval, and Cloud-Based RAG & IoT, smart home, healthcare, and personal assistance & ACM/IEEE International Conference on Connected Health: Applications, Systems and Engineering Technologies & C\\
P32 & Jaiswal (2025) & SAGESERVE: Optimizing LLM Serving on Cloud Data Centers with Forecast Aware Auto-Scaling & Network and service orchestration & Proceedings of the ACM on Measurement and Analysis of Computing Systems & J\\
P33 & Ouyang (2024) & Pluto and Charon: A Time and Memory Efficient Collaborative Edge AI Framework for Personal LLMs Fine-tuning & Industrial/IIoT/manufacturing and enterprise automation & International Conference on Parallel Processing & C\\
P34 & Nedisan Videsjorden (2025) & LUMEN: Enhancing IoT System Observability with Multi-Agent Large Language Models and Knowledge Graphs & IoT, smart home, healthcare, and personal assistance & ACM Transactions on Internet of Things & J\\
P35 & Jadhav (2025) & Evaluating the Efficacy of LLM-Based Reasoning for Multiobjective HPC Job Scheduling & Cross-cutting methods, security, and enabling substrates & SC'25 Workshops of the International Conference for High Performance Computing, Networking, Storage and Analysis & C\\
P36 & Saleh (2025) & MemIndex: Agentic Event-based Distributed Memory Management for Multi-agent Systems & Cross-cutting methods, security, and enabling substrates & ACM Transactions on Autonomous and Adaptive Systems & J\\
P37 & Du (2026) & TaskON: Task-Oriented Networking for Agentic AI & IoT, smart home, healthcare, and personal assistance & IEEE Network & J\\
P38 & Lin (2026) & An integrated framework featuring policy-governed agentic AI for closed-loop manufacturing control with multi-source sensor-MES-ERP & Industrial/IIoT/manufacturing and enterprise automation & The International Journal of Advanced Manufacturing Technology & J\\
P39 & Ji (2025) & Cloud-Edge Collaborative Service Architecture With Large-Tiny Models Based on Deep Reinforcement Learning & Vehicular, robotics, and mobile cyber-physical systems & IEEE Transactions on Cloud Computing & J\\
P40 & Zhang (2025) & Enhancing task offloading in vehicular networks: A multi-agent cloud-edge-device framework & Vehicular, robotics, and mobile cyber-physical systems & Vehicular Communications & J\\
P41 & Yao (2025) & Enhancing LLM QoS Through Cloud-Edge Collaboration: A Diffusion-Based Multi-Agent Reinforcement Learning Approach & Network and service orchestration & IEEE Transactions on Services Computing & J\\
P42 & Mujtaba (2025) & EdgeAIGuard: Agentic LLMs for Minor Protection in Digital Spaces & IoT, smart home, healthcare, and personal assistance & IEEE Internet of Things Journal & J\\
P43 & Luti (2025) & FUSCO: Distributed Agentic Data Acquisition and Fusion for Semantic Corner Case Detection & Vehicular, robotics, and mobile cyber-physical systems & IEEE Military Communications Conference (MILCOM) & C\\
P44 & Dettmers (2023) & QLoRA: Efficient Finetuning of Quantized LLMs & Cross-cutting methods, security, and enabling substrates & Advances in Neural Information Processing Systems & C\\
P45 & Piccialli (2025) & Federated and edge learning for large language models & Cross-cutting methods, security, and enabling substrates & Information Fusion & J\\
\end{longtable}
\end{scriptsize}

\section{Coding of the Primary Studies}
\label{app:consolidated-mapping}
\begin{small}
\setlength{\tabcolsep}{4pt}
\renewcommand{\arraystretch}{1.08}
\begin{longtable}{llllllll}
\caption{\BlueText{Coding of the 45 primary studies. Architecture: A1 edge-resident teams, A2 planner-worker orchestration, A3 hybrid local/cloud, A4 federated adaptation, A5 infrastructure. Deployment: D1 local-only execution, D2 edge-first with cloud escalation, D3 hierarchical orchestration, D4 federated adaptation, D5 state continuity and migration, D6 governance-gated control. Domain: M1 network and service orchestration, M2 IoT, smart home, healthcare, and personal assistance, M3 industrial and enterprise automation, M4 vehicular, robotics, and mobile systems, M5 cross-cutting methods and infrastructure. Evaluation families: E1 latency and QoS, E2 task success and quality, E3 privacy, safety, and trust, E4 cost and efficiency, E5 continuity and resilience. Setting: how the system was evaluated. Baseline: at least one baseline comparison reported. Artifact: public code or data found. A dash means that no code applies: P34 and P35 do not fix component placement, and P7 reports no experimental evaluation. $\dagger$ marks the two snowballed supporting records (Section \ref{sec:method}). Borderline settings are coded by the main experimental platform: P17 runs on a physical K3s cluster with simulated network dynamics, and P32 replays production traces.}}\label{tab:appendix-consolidated-mapping}\\
\hline
\textbf{ID} & \textbf{Arch.} & \textbf{Deploy.} & \textbf{Domain} & \textbf{Eval.} & \textbf{Setting} & \textbf{Baseline} & \textbf{Artifact}\\
\hline
\endfirsthead
\multicolumn{8}{l}{\emph{(continued from previous page)}}\\
\hline
\textbf{ID} & \textbf{Arch.} & \textbf{Deploy.} & \textbf{Domain} & \textbf{Eval.} & \textbf{Setting} & \textbf{Baseline} & \textbf{Artifact}\\
\hline
\endhead
\hline
\multicolumn{8}{r}{\emph{(continued on next page)}}\\
\endfoot
\hline
\endlastfoot
P1 & A1 & D1       & M2 & E2,E3      & Hardware & Yes & No\\
P2 & A2 & D3       & M1 & E1,E2      & Hardware & Yes & No\\
P3 & A1 & D1       & M2 & E2         & Hardware & Yes & No\\
P4 & A5 & D3       & M5 & E1,E4      & Simulation & Yes & No\\
P5 & A3 & D3       & M5 & E2,E4      & Hardware & Yes & No\\
P6 & A5 & D3       & M1 & E1,E4      & Hardware & Yes & No\\
P7 & A2 & D3       & M1 & --         & Conceptual & No & No\\
P8 & A3 & D2       & M2 & E2         & Hardware & Yes & No\\
P9 & A5 & D6       & M5 & E3,E4      & Simulation & Yes & No\\
P10 & A2 & D3       & M1 & E1         & Hardware & Yes & No\\
P11 & A5 & D1       & M2 & E2,E3      & Hardware & Yes & No\\
P12 & A2 & D3       & M1 & E1,E3      & Simulation & Yes & No\\
P13 & A2 & D3       & M3 & E2         & Simulation & Yes & No\\
P14 & A4 & D4       & M3 & E3,E4      & Hardware & Yes & No\\
P15 & A1 & D1       & M3 & E2         & Hardware & Yes & No\\
P16 & A2 & D3       & M1 & E1,E2      & Simulation & Yes & No\\
P17 & A4 & D3,D4    & M5 & E1,E4      & Hardware & Yes & Yes\\
P18 & A5 & D1       & M3 & E2         & Simulation & Yes & No\\
P19 & A1 & D3       & M4 & E2,E3      & Hardware & Yes & No\\
P20 & A3 & D2       & M1 & E2         & Simulation & Yes & No\\
P21 & A2 & D3       & M3 & E1,E2      & Production & Yes & No\\
P22 & A5 & D5       & M5 & E3,E5      & Hardware & Yes & No\\
P23 & A3 & D2       & M1 & E1         & Hardware & Yes & No\\
P24 & A5 & D5       & M4 & E5         & Hardware & Yes & No\\
P25 & A3 & D2       & M5 & E1,E4      & Hardware & Yes & Yes\\
P26 & A3 & D2       & M5 & E1,E2,E4   & Hardware & Yes & Yes\\
P27 & A5 & D1,D6    & M2 & E1,E3      & Hardware & Yes & No\\
P28 & A5 & D6       & M2 & E1,E2      & Hardware & Yes & No\\
P29 & A2 & D3       & M4 & E2         & Simulation & Yes & No\\
P30 & A3 & D2       & M2 & E1,E2,E4   & Hardware & Yes & No\\
P31 & A3 & D2,D6    & M2 & E3         & Hardware & No & No\\
P32 & A5 & D3       & M1 & E1,E4      & Production & Yes & Yes\\
P33 & A4 & D4       & M3 & E4         & Hardware & Yes & No\\
P34 & A1 & --       & M2 & E1,E2      & Hardware & Yes & No\\
P35 & A5 & --       & M5 & E1,E2,E4   & Simulation & Yes & Yes\\
P36 & A5 & D5       & M5 & E5         & Hardware & Yes & No\\
P37 & A5 & D3       & M2 & E1,E5      & Hardware & Yes & No\\
P38 & A2 & D6       & M3 & E2,E3      & Hardware & Yes & No\\
P39 & A3 & D2       & M4 & E4         & Simulation & Yes & No\\
P40 & A5 & D3       & M4 & E1,E4      & Simulation & Yes & No\\
P41 & A3 & D2       & M1 & E1,E4,E5   & Hardware & Yes & No\\
P42 & A1 & D1,D6    & M2 & E2,E3      & Hardware & Yes & No\\
P43 & A1 & D2,D6    & M4 & E1,E2,E4   & Hardware & Yes & No\\
P44$^\dagger$ & A4 & D4       & M5 & E4         & Simulation & Yes & Yes\\
P45$^\dagger$ & A4 & D4       & M5 & E4         & Conceptual & No & No\\
\end{longtable}
\end{small}

\section{Glossary of Terms and Acronyms}
\label{app:glossary}
\BlueText{
To aid readers who may not be familiar with the vocabulary of multi‑agent LLM systems and cloud‑edge‑IoT computing, this appendix collects concise definitions of the technical terms and acronyms used in the paper. The definitions are specific to the context of agentic AI, distributed systems, and continuum computing, and are ordered alphabetically.
\begin{small}
\setlength{\tabcolsep}{3pt}
\renewcommand{\arraystretch}{1.1}
\begin{longtable}[H]{p{0.27\linewidth} p{0.68\linewidth}}
\caption{Glossary of technical terms and acronyms used in this paper. Definitions are intended to be succinct and domain‑specific.}\label{tab:glossary}\\
\toprule
\textbf{Term or acronym} & \textbf{Definition}\\
\midrule
\endfirsthead
\multicolumn{2}{l}{\emph{(continued from previous page)}}\\
\toprule
\textbf{Term or acronym} & \textbf{Definition}\\
\midrule
\endhead
\midrule
\multicolumn{2}{r}{\emph{(continued on next page)}}\\
\endfoot
\bottomrule
\endlastfoot
Agent & An autonomous software component capable of perceiving inputs, reasoning, and taking actions.\\
Agentic AI & AI systems endowed with agency, enabling reasoning, tool use, decision making, and interaction within control loops.\\
Autoscaling & Dynamically adjusting computing resources (e.g., servers, GPUs) based on workload to optimise cost and performance.\\
Cloud & Centralised computing resources providing scalable processing and storage, typically remote from data sources.\\
Cloud‑edge‑IoT continuum & Distributed computing substrate spanning IoT devices, edge/MEC nodes and cloud services; also called the cloud‑to‑thing continuum.\\
Edge computing & Processing data close to its source on devices or gateways to reduce latency, save bandwidth, and preserve privacy.\\
Edge‑first escalation & Deployment strategy that performs local processing but invokes remote/cloud reasoning when tasks are complex or confidence is low.\\
Evaluation family & Category of measured outcomes, such as latency, task success, privacy, efficiency, or resilience.\\
Federated adaptation & Distributed learning or fine‑tuning where updates occur at clients or edge nodes without centralising raw data.\\
Federated learning & Collaborative machine learning paradigm in which models are trained across decentralised devices while keeping data local.\\
Fine‑tuning & Adapting a pre‑trained model to a specific task or domain using additional training data; may employ quantisation and low‑rank techniques.\\
Fog computing / MEC & Intermediate layer providing storage and computation between devices and cloud; MEC stands for Multi‑access Edge Computing.\\
Governance & Policies, permissions, and oversight that control data access, tool use, actuation, and audit logging for agentic systems.\\
Governance‑gated loop & Control loop that requires policy checks, authorisation, safety validation, or human approval before executing an action.\\
GPU‑hour & Unit of computational cost equal to one hour of GPU utilisation.\\
Hierarchical orchestration & Scheduling and placement of agent tasks across device, edge and cloud tiers, often managed by orchestrators or schedulers.\\
Hybrid local–small / cloud–large & Architecture using compact local models or modules for immediate tasks and larger cloud models for complex reasoning or fallback.\\
IIoT & Industrial Internet of Things; application of IoT technologies to industrial automation and manufacturing.\\
IoT & Internet of Things; network of physical devices embedded with sensors, actuators, and connectivity for data exchange and control.\\
Knowledge graph & Structured representation of entities and relationships used to support reasoning or retrieval augmentation.\\
Latency & Time delay between a request and its response; often measured as p50 (median) or p95 (95th percentile) latency.\\
LLM & Large Language Model; a neural network trained on vast corpora to interpret and generate human‑language text or instructions.\\
Local‑only execution & Deployment pattern in which reasoning, tool invocation, and actuation remain on the device or local edge node.\\
LoRA / QLoRA & Low‑Rank Adaptation (LoRA) and Quantised Low‑Rank Adaptation (QLoRA); techniques for parameter‑efficient fine‑tuning of LLMs, often using 4‑bit weights.\\
MCP & Model Context Protocol; an emerging standard for exchanging contextual data between models and tools to support tool‑augmented reasoning.\\
MEC & See Fog computing / MEC.\\
Memory state & Persistent context maintained by an agent across tasks or sessions, such as session memory or vector caches.\\
Migration & Moving an agent’s execution context or state from one node to another to maintain performance, connectivity, or resource utilisation.\\
Multi‑agent LLM system & Software system in which several LLM‑based agents cooperate through task decomposition, coordination, or communication.\\
Multi‑modal & Involving multiple data modalities (e.g., text, images, speech) in agent inputs or outputs.\\
Multi‑task & Handling different task types (e.g., classification, reasoning, planning) within one system.\\
Multi‑dialogue & Supporting multiple concurrent conversations between agents and users or other agents.\\
Offloading & Moving computation or tasks from a device to a more capable node (edge or cloud) to meet latency, energy, or cost objectives.\\
Ontology & Formal specification of concepts and relations in a domain, enabling semantic reasoning for agents.\\
PHI & Protected Health Information; sensitive medical data requiring stringent privacy protection and compliance.\\
Planner–worker orchestration & Architecture in which a planner agent decomposes tasks and assigns subtasks to worker agents according to dependencies.\\
Privacy‑enhancing technology & Technique (e.g., blurring, anonymisation, differential privacy) used to protect sensitive data before processing or sharing.\\
Quality‑of‑Service (QoS) & Performance characteristic (e.g., throughput, delay, reliability) required by a network or service.\\
RAG & Retrieval‑Augmented Generation; combining language model outputs with information retrieved from external sources.\\
ReAct & Reason‑and‑Act pattern for LLMs that interleaves reasoning steps with tool actions to improve correctness and interpretability.\\
Resilience & Ability of a system to maintain functionality under failures, network disruptions, mobility, or resource fluctuations.\\
Role‑specialised team & Architectural family where agents specialise in distinct roles (e.g., sensing, summarising, execution) and run near data sources.\\
Selective escalation & Triggering remote or cloud processing only under certain conditions, such as high task complexity or low confidence.\\
Service function chain (SFC) & Ordered sequence of network functions or tasks (e.g., firewalls, NATs); generalised here to sequences of AI tasks mapped across continuum layers.\\
SLM & Small Language Model; a compact model derived from or trained alongside an LLM for deployment on resource‑limited devices.\\
State continuity & Ensuring that an agent’s state is maintained across disconnections or migrations, often via session memory, vector caches, or state shadowing.\\
State shadowing & Creating a proxy or mirror of an agent’s state in a reachable location to handle network isolation or NAT traversal.\\
Substrate‑oriented control & Architectural family focusing on infrastructure concerns such as offloading, communication protocols, memory management, scheduling, and migration.\\
Task success & Measure of whether a task or subtask is completed correctly, often assessed by accuracy, precision, or success rate.\\
Time‑to‑first‑token (TTFT) & Responsiveness metric for generative models; time elapsed until the first output token is produced.\\
Vector cache & Local storage of vector‑embedded knowledge or context enabling efficient retrieval and reasoning at the edge.\\
\end{longtable}
\end{small}
}

\end{document}